\documentclass[twocolumn,aps,prb,epsf]{revtex4}

\usepackage{amsmath}
\usepackage{epsfig}
\usepackage{epsf}
\usepackage{array}

\newcommand\be{\begin{equation}}
\newcommand\ee{\end{equation}}

\begin{document}

\title{Phonon effects in molecular transistors} 


\author{A. Mitra, I. Aleiner, A. J. Millis}
\affiliation{Department of Physics, Columbia University, 538 W. 120th St, NY NY 10027}
\date{\today}

\begin{abstract}
A rate equation formalism is used to determine the effect of electron-phonon
coupling on the conductance of a molecule. Interplay between the phonon-induced
renormalization of the density of states on the quantum dot and the
phonon-induced renormalization of the dot-lead coupling is found to be
important. Whether or not the phonons are able to equilibrate in a time rapid
compared to the transit time of an electron through the dot is found to affect
the conductance. Observable signatures of phonon equilibration are presented.
\end{abstract}


\maketitle
\narrowtext
 
In recent years it has become possible to fabricate devices in which the
active element is a very small organic molecule \cite{Organicdevice}. Such a
device may be thought of as a 'quantum dot': a structure weakly coupled to the
macroscopic charge reservoirs ('leads') and small enough that the quantization
of the energly levels on the dot is important. Quantum dots fabricated using
conventional semiconductor technology have been extensively studied
experimentally \cite{Quantumdotrefs} and theoretically \cite{Quantumdottheory}.\
However, the use of small molecules may lead to new physics. In particular,
as electrons are added or removed from a small molecule the \textit{shape} of
the molecule is altered. The energies associated with this shape change are
not small, and the time scales may be comparable to those related to the flow
of electrons into and out of the molecule. Interesting recent data
indicate that these effects may lead to interesting structures in the
conductance spectra of the dot \cite{Ralph02,Zhitenev02,Park00}.

The shape change may be thought of as a coupling of electrons on the molecule
to phonon modes of the molecule. The subject of electron-phonon coupling in
quantum dots has received some theoretical attention
\cite{Glazman87,Wingreen89,Li95,Gogolin02,Alexandrov02,Balatsky02}, 
however the existing literature appears contradictory, and many papers
omit important physics or make apparently incorrect statements. For example,
the phonon-induced renormalization of the coupling between the molecule and
the leads is often neglected \cite{Balatsky02,Gogolin02,Alexandrov02}.
Further, some papers assert (incorrectly, we believe) that phonon sidebands
may be observed even in the linear-response conductance
\cite{Wingreen89,Balatsky02}. There is apparently no comprehensive discussion
of the relation between model parameters and the phonon-induced features of
the conductance. Further, a coupling of the dot to leads implies a mechanical
coupling between the dot and the external world which may allow the phonons on
the dot to relax to an equilibrium determined by the leads, instead of
following the electrons. The effect of this physics on the conductance has
apparently not been examined.

In this paper we  present a theoretical treatment of a simple model of
electron-phonon effects in a molecular quantum dot which addresses these
issues. Our treatment is valid in the limit (likely of greatest immediate
experimental relevance) in which the temperature is higher than the level
broadening induced by coupling the molecule to the leads, and is based on a
rate equation approach with coefficients derived via a golden-rule analysis.
We present results for average current as functions of gate and
source-drain voltage for two limiting cases: phonons uncoupled from the
external world and responding only to on-dot electrons (phonon equilibration
slow compared to dwell time of electron on dot), and phonons equilibrated to
the external world independent of the electron occupation (phonon
equilibration fast compared to dwell time of electron on molecule). 
Subsequent papers \cite{Mitra03b} will analyze the low-$T$ limit
and the equilibrium/nonequilibrium crossover.

For simplicity we consider here the case of a molecule with a single relevant
level, coupled to a single (Einstein) phonon mode and to two leads, which we
label as 'left' and 'right', described by the Hamiltonian
\begin{equation}
H=H_{dot}+H_{phonon}+H_{mix} \label{H}%
\end{equation}
with
\begin{eqnarray}
&H_{dot}&  = \varepsilon n_{d}+Un_{d}\left(  n_{d}-1\right)  +\lambda \omega_{ph}\left(
b^{\dag}+b\right)  n_{d}\label{Hdot}\\
&H_{phonon}&  = \omega_{ph}b^{\dag}b\label{Hphonon}\\
&H_{mix,el}&  =\sum_{a=L,R}V_{a}\sum_{p\sigma}\left(  c_{ap\sigma}^{\dag
}d_{\sigma}+H.c.\right)  \label{Hmix}%
\end{eqnarray}

\begin{figure}[b]
\epsfxsize=2.0in \centerline{\epsffile{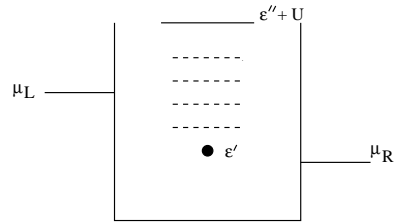}}
\vspace{0.5cm}
\caption{Energy level diagram. $\mu_{L,R}$
represent the chemical potential in the left and right leads respectively, while
$\epsilon^{\prime}$ represents the ground state of the singly occupied dot and 
the dashed lines indicate phonon excitations of the singly occupied dot.
The solid line indicates the energy $\epsilon^{\prime \prime} + U$ of the
doubly occupied dot. In this paper we take the $U \rightarrow \infty$ limit.}
\end{figure}
Here the number of electrons on the molecule, $n_{d}$, is given by
\begin{equation}
n_{d}=\sum_{\sigma}d_{\sigma}^{\dag}d_{\sigma}, \label{nd}%
\end{equation}
the parameter $U$ is the charging energy of the molecule, the parameter
$\varepsilon$ gives the energy of putting one electron into the dot, the $V$
represent the hybridization of the molecule and the leads, while the left and
right leads are characterized by chemical potentials $\mu_{L}$ and $\mu_{R}$
and densities of states $N_{L,R}$.\ We have defined the zero-phonon state to
be the ground state when $n_{d}=0$, and we neglect anharmonicity in the
lattice part of the Hamiltonian (such anharmonicity is of course induced by
the electron-phonon coupling and an intrinsic anharmonicity could easily be
added). 

Fig.~1 shows the energy level diagram corresponding to the problem we consider.
We choose as zero of energy the value $\epsilon^{\prime}$ corresponding to the ground
state of one electron on an isolated dot, and we consider the behaviour as the chemical
potentials of the left and right leads are varied. We specialize here to the 
large U  limit most likely to be relevant for small molecules, so restrict the
dot occupancy to zero or 1. In conventional language the source-drain 
voltage $V_{sd} = \frac{\mu_L-\mu_R}{e}$, while the gate voltage 
$V_g = \frac{\mu_L+\mu_R}{2e}$. One should include in the model 
a coupling between the phonons on
the molecule and phonons in the leads or elsewhere in the device. Such coupling
would provide a mechanism for the molecular phonons to equilibrate to the
leads, and should be included in a fully general treatment. Here we study only
the limit of uncoupled phonons (described by $H$ above) and the limit of fully
equilibrated phonons, which we obtain as described below.

We shall be interested in the current $I$ flowing between the leads, 
\begin{eqnarray}
I  &  =\frac{d}{dt}\sum_{p\sigma}\left\langle c_{Lp\sigma}^{\dag}c_{Lp\sigma
}-c_{Rp\sigma}^{\dag}c_{Rp\sigma}\right\rangle \label{I}
\end{eqnarray}
as a function of the two chemical potentials. 

The term $H_{mix}$ leads to broadening of the electron levels. The resulting
widths due to coupling to the left and right leads are $\Gamma_{L,R}%
=N_{L,R}V_{L,R}^{2}.$ We specialize here to the case in which temperature is
large compared to the level broadening: $\ \Gamma_{L,R}<<T.$ In this case, the
behavior may be described by rate equations describing transitions between
different states of the system \cite{Rate_Eqn}. \ A state of the system is
specified by a number of electrons on the dot, and a number of excited
phonons, and transitions involve changing the number of electrons or phonons
or both. \ Before writing the rate equations, it is convenient to perform a
standard \cite{Mahanch4} canonical transformation to eliminate the explicit
electron-phonon coupling in Eq.~\ref{H}. Defining 
$S=\lambda\left(  \sum_{\sigma}d_{\sigma}^{\dag}d_{\sigma}\right)  \left(
b^{\dag}-b\right)$
and transforming all operators $\mathit{O}$ via $e^{S}\mathit{O}e^{-S}$ leads
to a transformed Hamiltonian $H^{\prime}=H_{dot}^{\prime}+H_{mix}^{\prime}$
with
\begin{eqnarray}
H_{dot}^{\prime}  &  =\varepsilon^{\prime}n_{d}+\omega_{ph}a^{\dag}%
a+\tilde{U}n_{d}\left(  n_{d}-1\right) \label{Hprimedot}\\
H_{mix}^{\prime}  &  =\sum_{a=L,R}V_{a}\sum_{p\sigma}\left(  \widehat
{X}c_{ap\sigma}^{\dag}d_{\sigma}+H.c.\right)  \label{Hprimemix}%
\end{eqnarray}
where the transformed phonon operator $a=b-\lambda\sum_{\sigma}d_{\sigma
}^{\dag}d$ $_{\sigma}$, so the phonon ground state depends on the dot
occupancy. 
$\varepsilon^{\prime}=\varepsilon-\frac{g^{2}}{\omega_{ph}}$ 
is the 'polaron shift' in the energy for adding one electron to the
molecule and the  interaction parameter $U$ is also renormalized, but 
as we shall focus here on $U \rightarrow \infty$ we do not
write the renormalization explicitly here. The
crucial phonon renormalization of the electron-lead coupling is given by
\begin{equation}
\widehat{X}=\exp\left[  -\lambda\left(  a^{\dag}-a\right)  \right]
\label{Xdef}%
\end{equation}

After the transformation the state of the system is specified by the dot
occupancy $n$ and the number of phonons $q$ \textit{excited above the ground
state corresponding to the given occupancy }$n$ and we shall be concerned with
the probability $P_{q}^{n}$ that the system has $n$ electrons and $q$ excited
phonons. Transitions between states are determined by applying the golden rule
to $H_{mix}^{\prime}$. The resulting transition rate depends on the change in phonon
occupation number (it also depends on the electron distribution in the leads
but we write this dependence separately).

The transition rate involving hopping an electron from the dot to lead $a$ and
changing the phonon occupancy from $q$ (measured relative to the ground state
of $H_{dot}^{\prime}$ with occupancy $n$) to $q^{\prime}$ (measured relative
to the ground state of $H_{dot}^{\prime}$ with occupancy $n-1$) is equal to
the transition rate involving hopping an electron from the lead $a$ to the dot
and changing the phonon occupancy from $q$ (measured relative to the ground
state of $H_{dot}^{\prime}$ with occupancy $n-1$) to $q^{\prime}$ (measured
relative to the ground state of $H_{dot}^{\prime}$ with occupancy $n$) and is
\cite{Mahanch4}
\begin{equation}
\Gamma_{q^{\prime},q}^{a}=\Gamma_{a}\left\vert <q^{\prime}|X|q>\right\vert
^{2} \label{gqq'}%
\end{equation}
The matrix element can be computed by standard methods
\cite{Mahanch4}; its absolute value $|<q|X|q'>|^2 \equiv X_{qq'}^2 $ is symmetric 
under interchange of $q$ and $q'$  and is%
\begin{equation}
X_{q<q'}^2= \left\vert\sum_{k=0,q}\frac{\left(  -\lambda^{2}\right)  ^{k}\left(  q!q^{\prime
}!\right)  ^{1/2}\lambda^{|q-q^{\prime}|}e^{-\lambda^{2}/2}}{\left(  k\right)
!\left( (q-k\right)  !\left(  k+\left\vert q^{\prime
}-q\right\vert \right)  !}\right\vert ^{2} \label{X}%
\end{equation}

As interesting special cases, let us write the several lowest operators:%
\begin{mathletters}
\begin{eqnarray}
X_{0n}  &  = e^{-\lambda^{2}/2} \frac{\lambda^n}{\sqrt{n!}}\label{X0n} \\\
X_{11}  &  = \left(  1-\lambda^{2}\right)  e^{-\lambda^{2}/2}\label{X11}\\
X_{21}  &  =\sqrt{2}\lambda\left(  1-\frac{\lambda^{2}}{2}\right)
e^{-\lambda^{2}/2}\label{X21}\\
X_{22}  &  =\left(  1-2\lambda^{2}+\frac{\lambda^{4}}{2}\right)
e^{-\lambda^{2}/2} \label{X22}
\end{eqnarray}
\end{mathletters}
Observe that for certain values of $\lambda$ some of the matrix elements
vanish. This unusual behavior is an interference phenomenon, which is
slightly obscured by the notation. A state which has $q$ phonons excited
above {\it the ground state of the system with $n=0$ electrons} is 
a superposition (with varying sign) of many multiphonon states, when viewed in the
basis which diagonalizes the $n=1$ electron problem, and therefore the
transition described by $X_{qq'}$ is really a superposition of
many different transitions, which for some values of $\lambda$ may
destructively interfere. The phonon renormalization of the molecule-lead coupling
is apparently omitted, or treated in an average manner
which neglects the $q,q'$ dependent structure, in 
several recent papers \cite{Balatsky02,Gogolin02,Alexandrov02}.

We are now in a position to write rate equations, following e.g.~\cite{Rate_Eqn}. 
If the phonons are uncoupled from the leads we have ($f_{a}$ is the
fermi function for lead $a$).%
\begin{eqnarray}
\dot{P}_{q}^{n}  &  =\sum_{a,q^{\prime}}f_{a}\left(  \left(  q-q^{\prime
}\right)  \omega_{ph}+Un(n-1)\right) \Gamma_{q,q^{\prime}}^{a}%
P_{q^{\prime}}^{\left(  n-1\right)  } \\ \nonumber
&  + \left(  1-f_{a}\left(  \left(  q^{\prime}-q\right)
\omega_{ph}+Un(n+1)\right)  \right) \Gamma_{q,q^{\prime}}^{a}%
P_{q^{\prime}}^{\left(  n+1\right)  }\label{rateeqs}\\
&  -\left(  1-f_{a}\left(  \left(  q-q^{\prime}\right)
\omega_{ph}+Un(n-1)\right)  \right)  \Gamma_{q^{\prime},q}^{a}P_{q}%
^{n}\nonumber\\
&  -f_{a}\left(  \left(  q^{\prime}-q\right)  \omega
_{ph}+Un(n+1)\right)  \Gamma_{q^{\prime},q}^{a}P_{q}^{n}\nonumber
\end{eqnarray}

The net current flowing into the dot \ from lead $a$ is
\begin{eqnarray}
I^{a}&=\sum_{n,q,q^{\prime}}P_{q}^{n} f_{a}\left(  \left(  q^{\prime
}-q\right)  \omega_{ph}+Un(n+1)\right)  \Gamma_{q,q^{\prime}}^{a}\\ \nonumber
& - P_{q}^{n+1}\left(
1-f_{a}\left(  \left(  q-q^{\prime}\right)  \omega_{ph}+Un(n+1)\right)
\right)  \Gamma_{q^{\prime},q}^{a}  \label{Ia}
\end{eqnarray}
where the sum on n is from $0$ to $(d_{max}-1)$, $d_{max}$ being the  
maximum occupation of the dot.
Note that in steady state it follows from the rate equation that $I^{L}=-I^{R}$ which 
equals the total current across the dot.
\vspace{0.3cm}

The opposite limit, of phonons equilibrated to the leads (assumed here to be
at the same temperature) may be treated in steady state by forcing the 
probability distributions on the right hand side of Eq.~\ref{rateeqs}
to have the phonon-equilibrium form $P^{n}_{q} = P^n e^{-q\omega_{ph}/T}(1-e^{-w_{ph}/T})$.
In the $U \rightarrow \infty$ limit this ansatz implies that the 
probabilty $P^0$ that the dot is empty is given by,
\begin{equation}
P^0 = \frac{\sum_{a,q,q^{\prime}}\Gamma^a_{q,q{\prime}} e^{-q\omega_{ph}/T}
\overline{f}_{a,q,q^{\prime}}}
{\sum_{a,q,q^{\prime}}2 \Gamma^a_{q,q{\prime}} e^{-q^{\prime}\omega_{ph}/T} f_{a,q,q^{\prime}} 
+ \Gamma^a_{q,q^{\prime}} e^{-q\omega_{ph}/T}\overline{f}_{a,q,q^{\prime}}}
\end{equation}
where $\overline{f}_{a,q,q^{\prime}} = 
1-f_a\left(\left(q-q^{\prime}\right)\omega_{ph}\right)$,
$f_{a,q,q^{\prime}}= 1-\overline{f}_{a,q,q^{\prime}} $ while,  
$P^1 = 1 - P^0$.

In general for both equilibrated and unequilibrated cases the steady state rate equations
($\dot{P}_n = 0$)
constitute an eigensystem which we solve numerically. From these
solutions we have computed the current. Representative results are shown 
in Fig.~2 which plots the low-T current as a function of $V_{sd}$ for two gate
voltages: $V_g = 0 \,\, (\mu_L= - \mu_R$, upper panel) 
and $V_g=\frac{V_{sd}}{2} \,\, (\mu_R=0$, lower panel), for both
equilibrated and unequilibrated phonons. 
\begin{figure}
\epsfxsize=2.5in \centerline{\epsffile{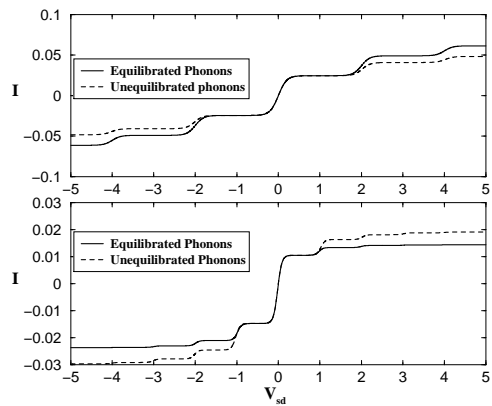}}
\caption{I-V for coupling constant $\lambda = 1.0$ as a function of $V_{sd}$ for 
$\omega_{ph}=1$ and $T=0.05$. Upper panel is for $V_g=0.0$, while lower panel
is for $V_{g} = V_{sd}/2, \mu_R = 0$.}
\end{figure}

Steps (broadened by T) in the 
current associated with ``phonon side-bands'' are observed when the 
\underline{source-drain voltage} passes through an integer multiple 
of the phonon frequency. However, in the opposite
'linear response' limit $V_{sd} \rightarrow 0$ (not shown),
as $V_g$ is varied
we find just one main step in the $I-V$-curve,
as $V_g$ passes through 0, and only very tiny structures
(vanishing as $e^{-\omega_{ph}/T}$, which is the probability of the dot being empty with
one phonon excited) when $V_g$ is a non-zero multiple of the 
phonon frequency. This result appears to differ from that stated in
Ref.~\cite{Balatsky02} who find phonon side bands as $V_g$ is swept at
$V_{sd} \rightarrow 0$. The difference occurs because Ref ~\cite{Balatsky02} 
neglected the fact that 
the phonon side-bands ``float'' {\it i.e.}, shift with the 
fermi level as  $V_g$  is changed.

Fig.~2 reveals on first sight an apparently surprising result: 
for symmetric bias ($V_g=0$)
and for the coupling considered, the current is larger for equilibrated
phonons than for the unequilibrated case, whereas for the strongly asymmetric
case ($\mu_R=0$), the opposite is true. This is surprising because one expects
that in the unequilibrated case the phonons arrange themselves so as to maximize
the current.
To gain more insight into this phenomenon we have calculated the dependence of the
ratio of currents for unequilibrated and equilibrated phonons 
on the coupling $\lambda$ for different degrees of bias asymmetry.
We find that except for $\mu_R=0.0$ (the most asymmetric case) 
a minimum in the ratio occurs
for a $\lambda \sim 1$. This behaviour may be traced back to Eq.~\ref{X11},\ref{X22}
which reveal that higher order ``diagonal'' (n phonon- n phonon) matrix
elements vanish for a $\lambda \sim 1$. 

The steps in current may be conveniently parametrized by
the height (or the area, as the width is simply proportional
to $T$) of the corresponding peaks $G_{max}$ in the differential
conductance $G=dI/dV$.
Ratios of peak heights (or areas) provide a convenient experimental measure of
whether the phonons are in equilibrium.
At low T, the equilibrium phonon distribution
corresponds to occupancy only of the $n=0$ phonon state, so the n-th side band
involves a transition from the 0 phonon to the n phonon state. Therefore the ratios
of the peak heights or areas are controlled by ratios of $|X_{n0}|^2$. 
In particular 
Eqns.~\ref{X0n},~\ref{Ia} imply that if $\mu_L = -\mu_R$ and $T\ll \omega_{ph}$,
\begin{equation}
\frac{G_{max}^n}{G_{max}^0}\Bigg\vert_{equil}
= \frac{|X_{n0}|^2}{2 |X_{00}|^2} = \frac{\lambda^{2n}}{2(n!)}
\end{equation}
Note that if $\mu_L \neq \mu_R$, then $\mu$ dependent changes in the 
occupation probabilities lead to additional, and not simply characterized
$n$ dependence.
\vspace{0.15cm}
\begin{figure}
\epsfxsize=2.5in \centerline{\epsffile{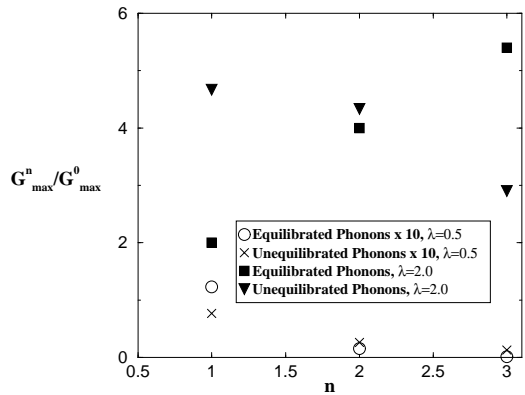}}
\vspace{-0.09cm}
\caption{
Ratio of peak heights for equilibrated and unequilibrated phonons and two 
different coupling strengths and $\mu_L = -\mu_R$. 
The points for $\lambda=0.5$ (open symbols) have been multiplied by 10.
}
\end{figure}

Deviations from this pattern imply non-equilibrium phonons. As illustration we display
in Fig.~3 $G_{max}$ values (normalized to the zero frequency peak) for equilibrium
and non-equilibrium phonons and a weak and strong electron phonon coupling. 
One sees that the non-equilibrium case the peak heights display a non-systematic
dependence on electron-phonon coupling and peak index, but that in
general measurements of the $n=1$ and $n=2$ peaks reveal the effect clearly.
\begin{figure}
\epsfxsize=2.5in \centerline{\epsffile{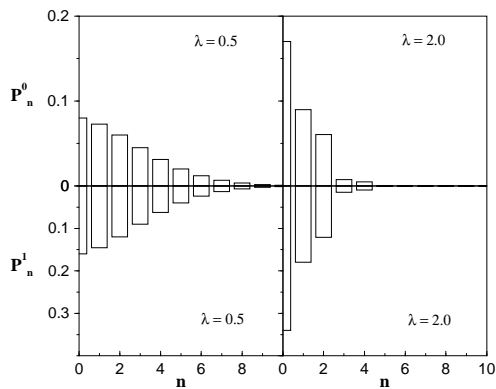}}
\caption{
Phonon probability distributions for two different electron-phonon coupling
constants calculated for $\mu_L = -\mu_R = 2\omega_{ph}$.
}
\end{figure}

It is also of interest to consider how far out of equilibrium the phonon distribution
may be driven. Fig.~4 shows the phonon occupation probabilities for  
weak and strong electron-phonon coupling and $V_g=0$. One sees
immediately that the phonon distribution function is farther from equilibrium 
for weak couplings than for strong couplings. We associate this effect to
the strong $\lambda$ dependence of operators $X_{0n}$, (Eq.~\ref{X0n}) which
allows the system at large $\lambda$ to ``jump down'' from a highly excited
state to one of low phonon occupancy. The deviation from equilibrium
is largest for $V_g=0$ for similar reasons.

In summary we have investigated a rate equation model for quantum dots with
strong electron-phonon coupling. We find important effects arising 
from the phonon renormalization of the dot-lead matrix element and from whether
or not the phonons are able to equilibrate.  Future papers will address
bistability \cite{Gogolin02,Alexandrov02} and also the low-T limit in
which the rate equation approach is not valid.

\textit{Acknowledgements: }\ AJM thanks L. Glazman for helpful conversations.
We acknowledge support from NSF-DMR-00081075 and Columbia University, and AJM
thanks the ESPCI for hospitality while part of this work was completed.


\end{document}